\documentclass[manuscript]{aastex}
\usepackage{longtable}
\usepackage{hyperref}
\usepackage{graphicx}
\usepackage{subfigure}
\usepackage{color}

\def\bc{\begin{center}}
\def\ec{\end{center}}
\def\be{\begin{eqnarray}}
\def\ee{\end{eqnarray}}

\slugcomment{Not to appear in Nonlearned J., 45.}

\shorttitle{Correlation Function of GRBs}
\shortauthors{Li et al.}

\begin{document}


\title{The Two-Point Correlation Function of Gamma-ray Bursts}



\author{Ming-Hua Li\altaffilmark{1}}
\affil{School of Physics and Engineering, Sun Yat-Sen University, Guangzhou 510275, China}

\author{Hai-Nan Lin\altaffilmark{2}}
\affil{Institute of High Energy Physics, Chinese Academy of Sciences, Beijing 100049, China}


\altaffiltext{1}{limh@ihep.ac.cn}
\altaffiltext{2}{linhn@ihep.ac.cn}


\begin{abstract}
In this paper, we examine the spacial distribution of gamma-ray bursts (GRBs) using a sample of 373 objects. We subdivide the GRB data into two redshift intervals over the redshift range $0<z< 6.7$. We measure the two-point correlation function (2PCF), $\xi(r)$ of the GRBs. In determining the separation distance of the GRB pairs, we consider two representative cosmological models: a cold dark matter universe plus a cosmological constant $\Lambda$, with $(\Omega_{{\rm m}}, \Omega_{{\rm \Lambda}})=(0.28,0.72)$ and an Einstein-de Sitter (EdS) universe, with $(\Omega_{{\rm m}}, \Omega_{{\rm \Lambda}})=(1,0)$. We find a $z$-decreasing correlation of the GRB distribution, which is in agreement with the predictions of the current structure formation theory. We fit a power-law model $\xi(r)=(r/r_0)^{-\gamma}$ to the measured $\xi(r)$ and obtain an amplitude and slope of $r_0= 1235.2 \pm 342.6~h^{-1}$ Mpc and $\gamma = 0.80\pm 0.19 $ ($1\sigma$ confidence level) over the scales $r=200$ to $10^4~h^{-1}$ Mpc. Our result provide a supplement to the measurement of matter correlation on large scales, while the matter distribution below $200~h^{-1}$ Mpc is usually described by the correlation function of galaxies. \end{abstract}




\section{Introduction}
Redshift surveys of galaxies have long been carried out to study the large-scale structure (LSS) of our Universe. The two-point correlation function (2PCF) is a statistic that can be easily determined from a well-measured galaxy sample. Theoretical predictions of the 2PCF can also be made from different dark matter models and structure formation scenarios. Thus the 2PCF of galaxies or quasi-stellar objects (QSOs, or quasars) has been used as an important statistic to distinguish different theoretical models \citep{Kundic1997, Eisenstein2005, Ross2007}.

Gamma-ray bursts (GRBs) are the most energetic events known to occur in the Universe. They are often associated with the death of massive stars and have a redshift up to $z\sim 8$. They are robust indicators of the matter-dense part of the intermediate- or even the high-redshift Universe. Their spacial distribution helps explore the LSS of the Universe. Although recent works suggested that GRBs have an anisotropic distribution in the sky \citep{Cai2013, Chang2014, Wang2014}, most studies have demonstrated that GRBs are distributed isotropically \citep{Briggs1996, Meszaros2000, Magliocchetti2003, Vavrek2008}, the latter of which is demanded by the cosmological principle. 

Besides its isotropic distribution in the sky, a homogeneous distribution of the GRBs is also expected. The discovery of a possible structure in the GRB sky distribution has been recently reported \citep{Horvath2014}. It has a redshift $z\simeq 2$ (at a distance of approximately ten billion light years away) and its size is about $2000$ to $3000$ Mpc. This excess clustering of GRBs has a statistical significance of 3$\sigma$ confidence level (c.l.) and therefore cannot be attributed to the sampling biases at this redshift. In the current structure formation theory, structures in today's Universe like galaxies and galaxy clusters etc. all stem from the primordial Gaussian random phase fluctuations of the mass density field. Given the finite time from the end of the cosmic inflation to the present, the evolution of the non-linear structures in our Universe is limited. They should not be larger than the scale $r_c$ on which the density contrast becomes $\delta_c \simeq 1$ at present as predicted by the linear growth theory of density perturbations\footnote{At the present time, on a scale smaller than $r_c$, the growth of density contrast would have become nonlinear.} \citep{Labini2008}. Since GRBs are potential tracers of normal matter, the discovery of this GRB structure (if confirmed by future investigations) casts new shadows on the cosmological principle as well as the current structure formation theory. 
In fact, a number of interesting results have been reported on the homogeneity scale of the galaxy and quasar distributions \citep{Croom2005, Yadav2005, Sarkar2009, Scrimgeour2012}.  To do a similar clustering analysis on the distribution of GRBs,
the correlation function of the GRB distribution has to be first measured.

In this work, we measure the real-space two-point correlation function $\xi(r)$ of GRBs. We use the catalogue presented by \citet{GRBs}. It contains 373 objects over a redshift range of $0<z<6.7$ by September 3rd, 2014. We subdivide the samples into two redshift regimes: $z<2$ and $z>2$. The 2PCF of each regime is calculated. We fit a power law to both of the measured $\xi(r)$. In numerical analysis, we prefer the $\Lambda$CDM (cold dark matter plus a cosmological constant) cosmological model, with $(\Omega_{{\rm m}}, \Omega_{{\rm \Lambda}})=(0.28,0.72)$ over the Einstein-de Sitter (EdS) Universe, with $(\Omega_{{\rm m}}, \Omega_{{\rm \Lambda}})=(1,0)$. 

The rest of the paper is organized as follow: in Section 2, we introduce the methodology of the estimation of the correlation function, including the estimators, the biases, and the error calculations. In Section 3.1, we describe the data we use and the auxiliary random sample. In Section 3.2, we plot the measured $\xi(r)$ for the GRBs. In Section 3.3, we fit a power law to the measured $\xi(r)$ and give the best-fit values of amplitude and slope. Conclusion and discussions are presented in Section 4.

\section{Estimation of the correlation function}
\subsection{The two-point correlation function and the estimators}
The 2PCF, $\xi(x)$, is defined by the probability of finding an object in the spacial volume $dV_1$ with another object in $dV_2$ that is separated by a distance $x$, i.e. \citep{Peebles1980}
\be
dP={\bar n}^2[1+\xi(x)]dV_1dV_2.
\ee
${\bar n}$ is the mean number density of the objects. For statistical estimation of $\xi(x)$, an auxiliary random sample of $N_R$ points must be generated in a window $W$. A window $W$ is a three-dimensional space of volume $V$ equivalent to that on which the observation is made. Following \citet{Kerscher2000}, we define the pair count with a unitary function $F(\mathbf{x},\mathbf{y})$:
\be
N_{DR}(r)=\sum_{\mathbf{x}\in D}\sum_{\mathbf{y}\in R}F(\mathbf{x},\mathbf{y}).
\label{DR}
\ee
The summation runs over all the coordinates of objects (represented by $\mathbf{x}$ and $\mathbf{y}$) in the observed data set $D$ and the random sample $R$ in the window $W$. The value of the function $F(\mathbf{x},\mathbf{y})$ equals 1 when the separation of the two objects is within the distance $d(\mathbf{x},\mathbf{y})\in [r-\Delta r/2, r+\Delta r/2]$ or otherwise equals 0. $d(\mathbf{x},\mathbf{y})$ is the comoving distance between the two objects and $\Delta r$ is the bin width being used in the statistical estimation of $\xi(r)$. With these preparations, we can determine the 2PCF from the observed data. 
 
Several estimators of $\xi(r)$ are popular. One was presented by \citet{DP1983}, i.e. the DP estimator,
\be
\hat{\xi}_{{\rm DP}}(r)=\frac{DD(r)}{DR(r)}-1.
\label{DP83}
\ee
$DD(r)$ is the normalized number of all pairs of GRBs in the observed data set and is given as $DD(r)\equiv N_{DD}(r)/[N_D (N_D-1)]$, where $N_D$ is the total number of GRBs in the data set. $DR(r)\equiv N_{DR}(r)/(N_D N_R)$ is the normalized number of pairs between the data and the random sample with separation of $d(\mathbf{x},\mathbf{y})\in [r-\Delta r/2, r+\Delta r/2]$. $N_{DD}(r)$ is defined in a similar way as $N_{DR}(r)$ in equation (\ref{DR}), while `$D$' and `$R$' refer respectively to the observed data set of GRBs and the auxiliary random catalogue.
Two other widely used estimators are the Landy-Szalay (LS) estimator $\hat{\xi}_{{\rm LS}}(r)$ \citep{LS1993} and the Hamilton estimator $\hat{\xi}_{{\rm Ham}}(r)$ \citep{Hamilton1993}:
\be
&&\hat{\xi}_{{\rm LS}}(r)=\frac{DD(r)- 2DR(r)+RR(r)}{RR(r)}, \\
\label{xiLS}
&&\hat{\xi}_{{\rm Ham}}(r)=\frac{DD(r) RR(r)}{[DR(r)]^2}-1.
\label{xiH}
\ee
$RR(r)\equiv N_{RR}(r)/[N_R(N_R-1)]$ refers to the normalized number of pairs with the separation mentioned above in the random sample. $N_{RR}(r)$ is defined in a similar way as $N_{DR}(r)$ in equation (\ref{DR}).

\subsection{The biases}
A relevant problem for estimating the 2PCF is that there might be systematic biases and  stochastic noise which perturbs any real determination of the 2PCF, especially on small-amplitude values of $\xi(r)$. At small distances, the estimators mentioned above have very similar performances. However, on large scales, they are not totally equivalent and some of them could be biased. 

Generally speaking, for a valid statistical estimator of the quantity $X$, the sample average $\overline{X}$ in a finite volume $V$, i.e. $\overline{X(V)}$, must satisfy \citep{Gabrielli2004}
\be
\lim_{V\rightarrow \infty}\overline{X(V)}= \langle X\rangle ,
\label{validestimator}
\ee
where $\langle X\rangle$ is the ensemble average. For a finite sample volume $V$, there is a systematic bias in the estimator $\hat{X}$. 
An unbiased estimator is the one that satisfies $\overline{X(V)}= \langle X\rangle$. 

The LS estimator is generally biased as the DP and Hamilton estimators. There are also other two biased estimators that are commonly used, i.e. the natural estimator and the Hewett estimator. 
Numerical tests on artificial distributions have been done to study biases in these estimators \citep{Pons-Borderia1999, Labini2008}. It has been found that on large scales, the LS estimator and the Hamilton estimator significantly outperform the rest \citep{Pons-Borderia1999}. 
The LS estimator has an indistinguishable performance as the Hamilton estimator but is less sensitive to the size of the random sample, $N_R$. Thus it is more preferable from a practical point of view.

In this work, we focus on the LS estimator of $\xi(r)$, while the DP and Hamilton estimator are also used for comparison.
A detailed comparison of these estimators can be found in \citet{Kerscher2000}, while alternative estimators (such as the full-shell estimator and the geometric estimator, etc.) were proposed in \citet{Kerscher2000} and \citet{Labini2008}.

\subsection{The errors}
A limited sample would result in a large likely error or variance. A proper way of estimating errors could reduce this effect to the least. 
There are several ways\footnote{One can refer to \citet{Hamilton1993} and see the references therein. A so-called `cox process' has also been given in \citet{Pons-Borderia1999}.}  to determine the measured errors of the 2PCF of GRBs. Three most common ones are the Poisson estimate,   the `field-to-field' (FtF) error, and the jackknife estimate. The {\em Poisson} estimate of the errors of $\xi(r)$ is given as:
\be
\sigma_{{\rm Poi}}(r)=\frac{1+\xi(r)}{DD(r)}.
\label{poisson}
\ee

The second method is the {\em `FtF'} method. In this method, the whole sample is divided into $N_b$ subsamples. The 2PCF of each subsample is calculated, i.e. $\xi_i(r)$ with $i=1,2, ... , N_b$.
The error is calculated by
\be
\sigma^2_{{\rm FtF}}(r)=\frac{1}{N_b-1}\sum_{i=1}^{N_b}\frac{DR_i(r)}{DR(r)}[\xi_i(r)-\xi(r)]^2.
\label{ftf}
\ee
$\xi(r)$ is the estimate of the 2PCF on the entire sample. For our studies, the entire sample of GRBs, which has a redshift range of $z\in [0, 6.7]$, is divided into seven redshift bins with the interval of 1. Thus there is $N_b=7$. Details of the subsamples are given in Table \ref{table1}.

The third way to estimate the errors of $\xi(r)$ is called the {\em jackknife} estimate. It is given as
\be
\sigma^2_{{\rm Jack}}(r)=\sum_{i^\prime =1}^{N^\prime}\frac{DR_{i^\prime} (r)}{DR(r)}[\xi_{i^\prime} (r)-\xi(r)]^2.
\label{jackknife}
\ee
Like the `FtF' method, the entire sample is divided into $N^\prime$ subsamples. $\xi_{i^\prime} (r)$ denotes the estimate of the 2PCF on all of the $(N^\prime -1)$ subsamples except the $i$-th one.

On small scales $r \lesssim 10h^{-1}$ Mpc ($h$ is the Hubble constant in units of $100$ km s$^{-1}$ Mpc$^{-1}$), all these errors have comparable magnitudes, while above this scale (up to $r \simeq 100h^{-1}$ Mpc), the jackknife and `FtF' errors are considerably larger than the Poisson estimate \citep{Ross2007}. In addition, on small scales, the jackknife method gives smallest fluctuations of the three and is thus more efficient \citep{Scranton2002, Zehavi2002}. So in this work, we make use of the jackknife method to estimate the errors of the measured 2PCF. Like the `FtF' error, we divide the whole GRB sample into 7 subsamples as well as 7 redshift bins. Each subsample corresponds to one of these redshift bins. Thus there is $N^\prime=7$.

\section{Data and Results}
\subsection{The GRB data and the random catalogue}
We use the GRB catalogue\footnote{$~$See \url{http://www.mpe.mpg.de/~jcg/grbgen.html}} presented by J. Greiner to determine the 2PCF. It is a subjective collection of GRBs that are detected by a number of satellites and programs, i.e. BATSE, RXTE, BeppoSAX, HETE-2, Interplanetary Gamma-Ray Burst Timing Network (IPN), INTEGRAL, Swift, AGILE, and the Fermi Gamma-Ray Observatory. 
The catalogue contains thousands of objects and is updated almost everyday. We use the data that were released before September 3rd in 2014. Among those 398 objects which have the redshift measured, only 375 of them are well determined. 300 (about 80$\%$) of them come from one source, the Swift satellite. The rest of the samples come from Fermi Gamma-Ray Observatory, HETE-2, and other telescopes. For the others only an upper limit is provided. We subdivide these 375 GRBs into 10 redshift intervals with effective redshifts from $z =0$ to $9.2$. In the redshift bins with $z>7$ in the catalogue, there are only two GRBs: the GRB 090423 ($z=8.26$) and the GRB 090429B ($z=9.2$). They are omitted in our studies since they have little statistical significance. 
 
\begin{figure}
\includegraphics[scale=0.65]{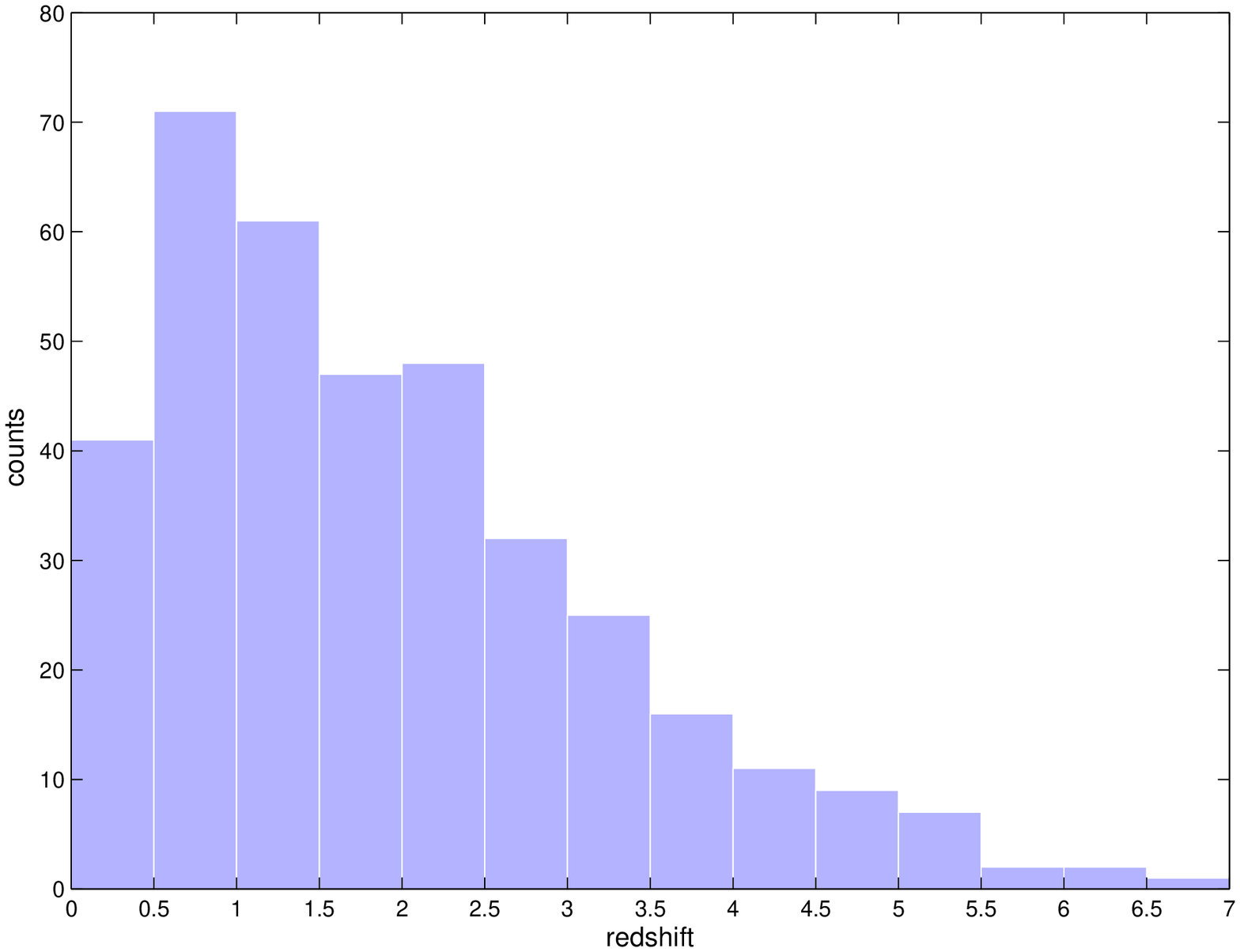}
\caption{The redshift distribution of the 373 GRB data. The $y$-axis denotes the number of objects in each redshift bin.}
\label{fig1}
\end{figure}

Therefore, we base our research on these $N_D=373$ GRBs that have redshifts and angular positions well determined. They cover a redshift range of $0<z<6.7$ and we use $N^\prime=7$ for the estimates of $\xi(r)$ and the jackknife errors. 
Robustness of the estimators introduced in the last section has been well demonstrated for a sample of hundreds of objects \citep{Shaver1984,Shaver1987,Shanks1987,AKS1988,IS1988,MF1993,SB1994,CS1996,Pons-Borderia1999,Kerscher2000}.
The GRB data we use are listed in Table 3 which is publicly available online\footnote{ The GRBs data we use is presented in a text format, which is publicly available online.}. The redshift distribution of the data are presented in Table \ref{table1} and Figure \ref{fig1}.
The density of random points used is 20 times the density of GRBs. We use bin widths of $\Delta {\rm log}(s)=2h^{-1}$ Mpc.

\subsection{The real-space two-point correlation function $\xi(r)$}
The measured values of 2PCF heavily rely on the assumed cosmology in that the comoving separation of the GRBs are model dependent. In this work, two cosmological models are considered: 1) the Lambda-cold dark matter ($\Lambda$CDM) cosmological model which uses the parameters determined from the nine-year Wilkinson Microwave Anisotropy Probe (WMAP) observations \citep{Bennett2013, Hinshaw2013} in conjunction with other observations\footnote{ Other observations include the high-$\ell$ cosmic microwave background (CMB) power spectrum data \citep{Fowler2010, Das2011, Keisler2011, Reichardt2012}, the baryonic acoustic oscillation (BAO) data \citep{Beutler2011, Padmanabhan2012, Anderson2012, Blake2012}, and a new $H_0$ measurement \citep{Riess2011}.}, with $(\Omega_{{\rm m}}, \Omega_{{\rm \Lambda}})=(0.28,0.72)$; 2) an Einstein-de Sitter (EdS) cosmological model, with $(\Omega_{{\rm m}}, \Omega_{{\rm \Lambda}})=(1,0)$.

\begin{figure}
\subfigure[~\textsf{$(\Omega_{{\rm m}}, \Omega_{{\rm \Lambda}})=(0.28,0.72)$}] { \label{fig1a}
\scalebox{0.52}[0.52]{\includegraphics{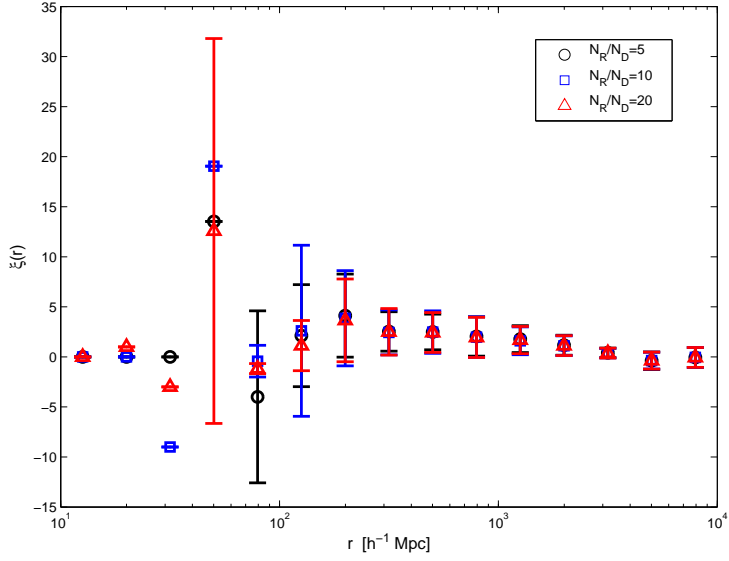}}
}
\subfigure[~\textsf{$(\Omega_{{\rm m}}, \Omega_{{\rm \Lambda}})=(1,0)$}] { \label{fig1b}
\scalebox{0.52}[0.52]{\includegraphics{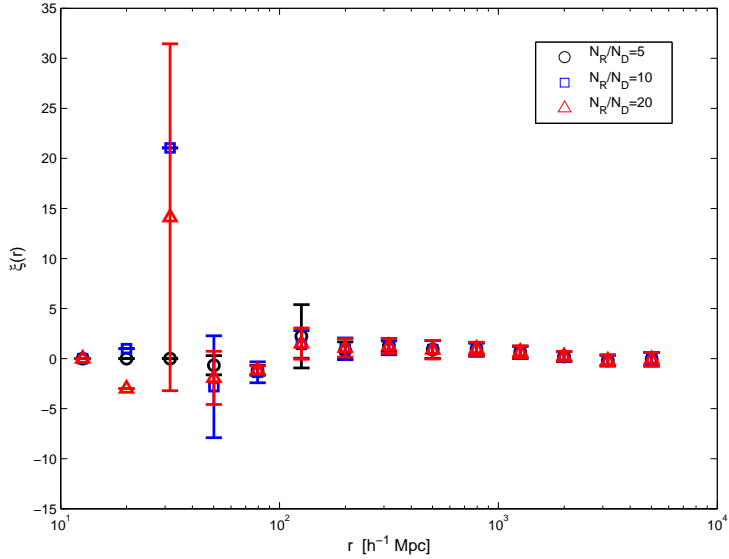}}
}
\caption{The real-space correlation function, $\xi(r)$ of the GRBs for different number of objects in the auxiliary random catalogue.
The results are presented in both the $\Lambda$CDM model, $(\Omega_{{\rm m}}, \Omega_{{\rm \Lambda}})=(0.28,0.72)$ and the EdS Universe, $(\Omega_{{\rm m}}, \Omega_{{\rm \Lambda}})=(1,0)$. Circles, squares, and triangles respectively represent the results obtained from that the random catalogues are 5, 10, and 20 times the density of GRB data. Error bars are calculated from the jackknife method by equation (\ref{jackknife}) with the $N^\prime =7$.}
\label{fig2}
\end{figure}

The 2PCFs of the GRB data for different densities of the random points are presented in Figure \ref{fig2}. We plot these to study the potential impacts of the statistical noise on the results. They cover a distance scale up to $\sim 800h^{-1}$ Mpc.
Most GRB pairs in the data have a separation distance over 100$h^{-1}$ Mpc. The current structure formation theory predicts that on such a scale, the clustering of matter remains in the linear regime even today \citep{Springel2005, Eisenstein2007}. The redshift space distortions\footnote{ See Section 9.4 in \citet{Dodelson2008}.} due to the small-scale peculiar velocities of the objects and the redshift variances are also minimal on this scale \citep{Ross2007}. Thus the difference between the redshift-space and the real-space correlation functions on such large scales could be negligible. 

Two other comments on Figure \ref{fig2} are necessary. The first one is that the $\xi(r)$ of GRBs is quite scattered on scales below $100h^{-1}$ Mpc. The poor performance of the estimates of $\xi(r)$ and its errors on such a scale is due to the lack of data for $r<100h^{-1}$ Mpc. Most GRBs have been found to have a separation with each other above $100h^{-1}$ Mpc. More observations are needed to improve the performance of the analysis on $r<100h^{-1}$ Mpc. On the scale $r>100h^{-1}$ Mpc, the results don't show much difference for varying the objects density in the random catalogue. Thus they are more reliable and can be used to find the possible power law fit of $\xi(r)$.

The second one is that it shows a `bump' at $r=200h^{-1}$ Mpc in the measured $\xi(r)$ for the GRB sample. For an EdS Universe, it is located at $r\simeq 100 h^{-1}$ Mpc. For different number densities of the random points, the bump exists, implying that it cannot be attributed to statistical noise. In cosmology, the temperature drop of the Universe at about 380,000 years after the big bang would cause a sudden decline of sound speed in the matter fluid. This would leave the oscillations in it become frozen \citep{Martinez2009}. For the $\Lambda$CDM model with $(\Omega_{{\rm m}}, \Omega_{{\rm \Lambda}})=(0.28,0.72)$, the signal manifests as a peak at about $100 h^{-1}$ Mpc in the correlation function for the galaxy distribution \citep{Eisenstein2005}, which is about $100h^{-1}$ Mpc away from the `bump' we discovered. However, since the size of the GRB sample we use is not sufficiently large, whether these two have any physical connections or not is still subject to future investigations. 

\begin{figure}
\includegraphics[scale=0.55]{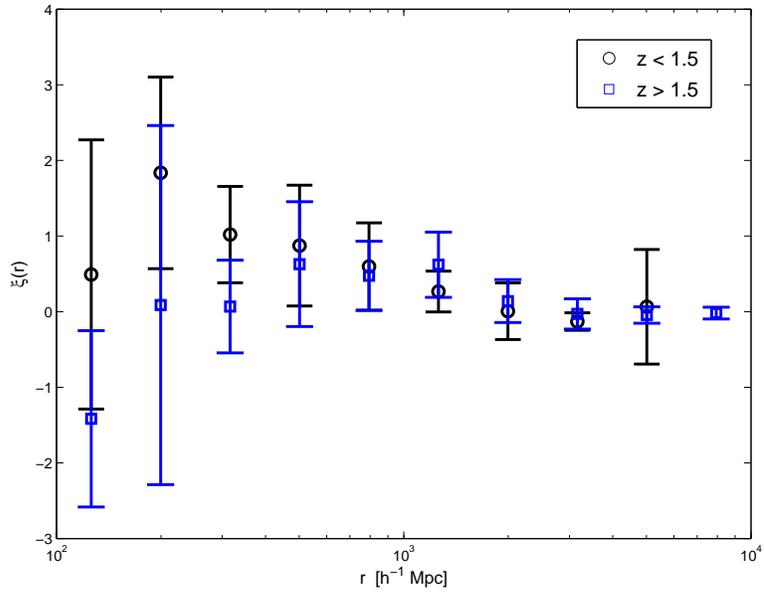}
\caption{The redshift evolution of the measured $\xi(r)$ for GRBs in the $\Lambda$CDM model with $(\Omega_{{\rm m}}, \Omega_{{\rm \Lambda}})=(0.28,0.72)$. The $\xi(r)$ for $z<1.5$ (black circles) and for $z>1.5$ (blue squares) with $1\sigma$ jackknife error bars are plotted.}
\label{fig3}
\end{figure}

To study the redshift evolution of the 2PCF for GRBs, the entire GRB samples are subdived into two groups with $z<1.5$ (173 objects) and $z>1.5$ (200 objects). The 2PCFs are calculated for each group for the $\Lambda$CDM model with $(\Omega_{{\rm m}}, \Omega_{{\rm \Lambda}})=(0.28,0.72)$ and are plotted in Figure \ref{fig3} for comparison. From Figure \ref{fig3}, it is found that on the scales $r <10^3 h^{-1}$ Mpc, the low-$z$ group of GRB samples have a higher correlation amplitude than the high-$z$ group. This is compatible with the current structure formation theory predicting that the precursors of GRBs with low redshifts would have more time to grow and thus would become more correlated with each other. The differences become small and indistinguishable on scales $r\gtrsim 5\times 10^2h^{-1}$ Mpc. In fact, some previous studies have suggested a $z$-increasing correlation behavior in quasars and halos \citep{Kundic1997, Franca1998, Gao2005}. \citet{Kundic1997} had reported that the amplitude of the quasar correlation function of the high-redshift samples ($z>2$) was significantly higher than that of the low-redshift sample ($z<2$). As shown in Figure \ref{fig3}, this is not the case for the GRB samples. More discussions are presented in Section \ref{conclusion}.

\subsection{A power law fit to $\xi(r)$}
A power law of the form 
\be
\xi(r)=\left(\frac{r}{r_0}\right)^{-\gamma}
\label{powerlaw}
\ee
has been respectively fitted to the 2PCFs of galaxies and galaxy clusters for a cold dark matter (CDM) Universe $(\Omega_{{\rm m}}=1)$, using a least-$\chi^2$ technique. For the case of galaxies, on scales $r\leq10h^{-1}$ Mpc, the correlation length $r_0$ has a best-fit value of $3.76\leq r_0 \leq 7.3~h^{-1}$ Mpc, with the exponent index $1.5\leq \gamma \leq 1.8$ \citep{DP1983, Maddox1990, Hermit1996, Zehavi2002, Zehavi2004}. The values vary somewhat from literature to literature depending on the galaxy sample utilized, the estimator used, and the weighting scheme employed, etc. For the case of galaxy clusters, the 2PCF also follows a power law, i.e $\xi_{{\rm cc}}(r)=(r/r_0)^{-1.8}$, with $13 \leq r_0 \leq 25~h^{-1}$ Mpc \citep{Bahcall1988, Peacock1992, Postman1992, Nichol1992, Dalton1992, Dalton1994}.

\begin{figure}
\subfigure[~\textsf{$(\Omega_{{\rm m}}, \Omega_{{\rm \Lambda}})=(0.28,0.72)$}] { \label{fig1a}
\scalebox{0.4}[0.4]{\includegraphics{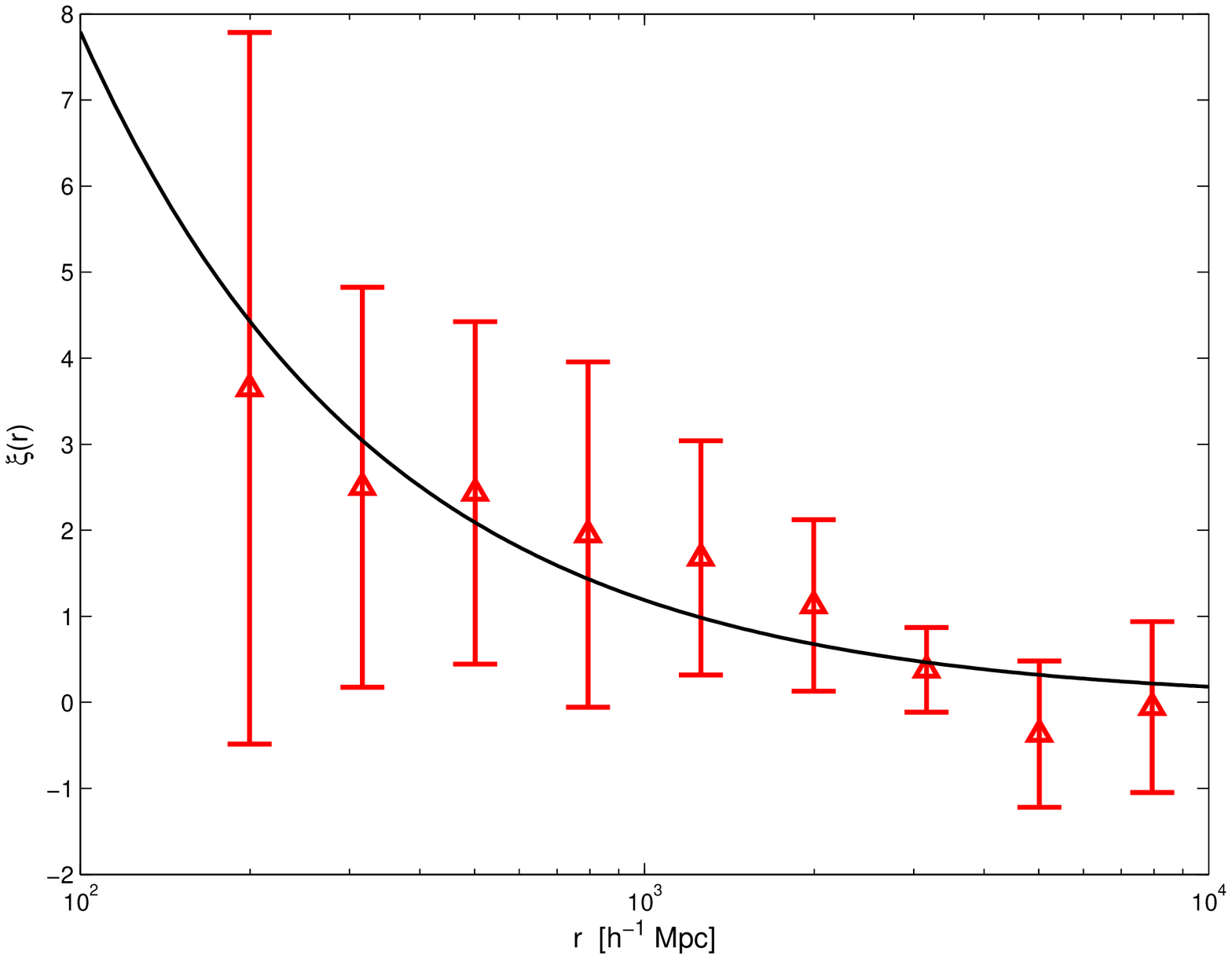}}
}
\subfigure[~\textsf{$(\Omega_{{\rm m}}, \Omega_{{\rm \Lambda}})=(1,0)$}] { \label{fig1b}
\scalebox{0.4}[0.4]{\includegraphics{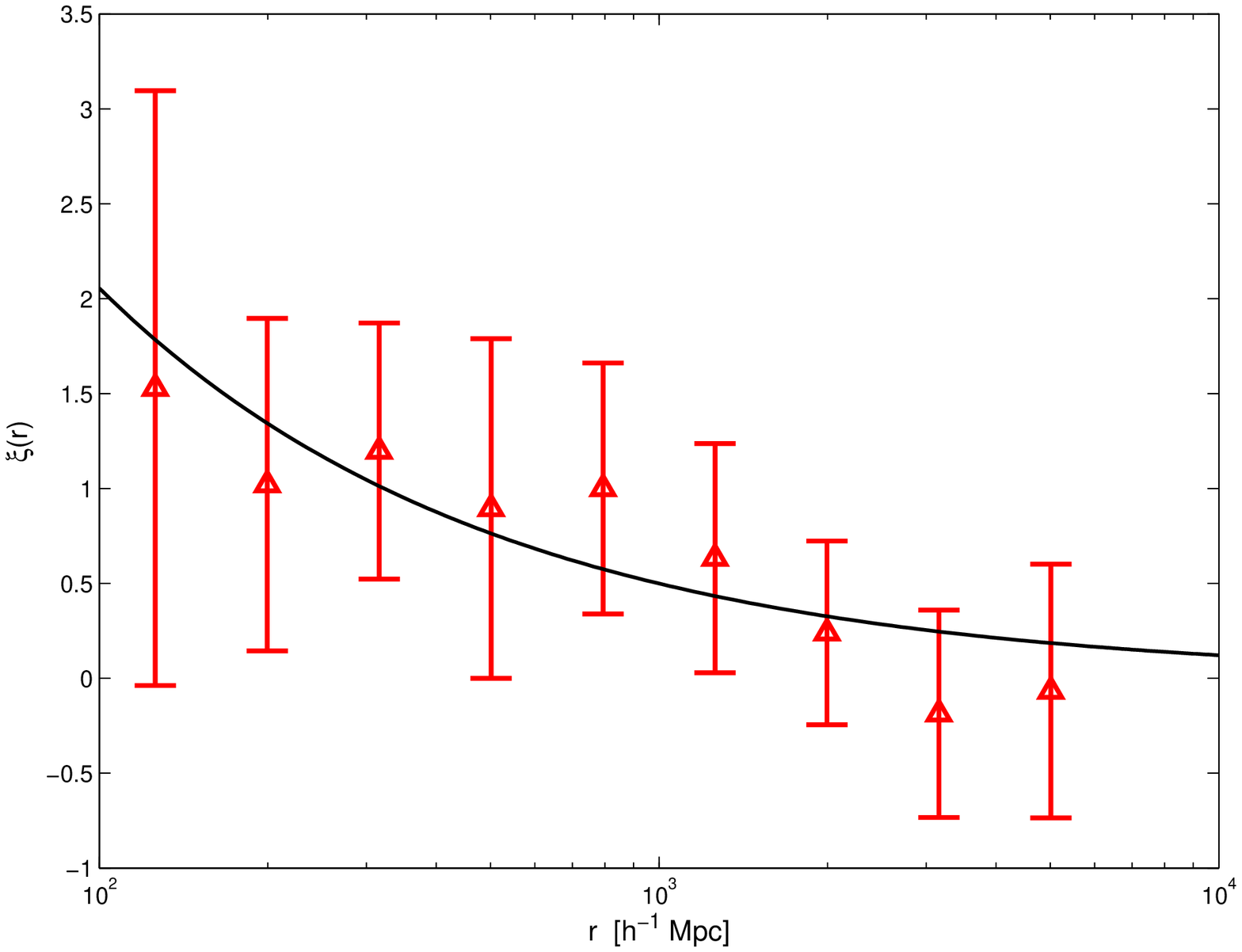}}
}
\caption{The best-fit power law of the measured $\xi(r)$ for the GRBs from \citet{GRBs}. (a) For the $\Lambda$CDM universe with $(\Omega_{{\rm m}}, \Omega_{{\rm \Lambda}})=(0.28,0.72)$. The best-fit values of the correlation length $r_0$ and the slope $\gamma$ are $r_0=1235.2 \pm 342.6$ $h^{-1}$ Mpc and $\gamma =0.80\pm 0.19$ ($1\sigma$ confidence level), with ${\bar \chi^2}_{{\rm min}}=0.19$. (b) For the EdS Universe with $(\Omega_{{\rm m}}, \Omega_{{\rm \Lambda}})=(1,0)$. The best-fit parameters are $r_0=322.4\pm92.3$ $h^{-1}$ Mpc and $\gamma =0.62\pm0.20$ ($1\sigma$ confidence level), with ${\bar \chi^2}_{{\rm min}}=0.21$. In both figures, triangles represent the measured $\xi(r)$ obtained from that the density of random points is 20 times of that of the GRB data. The $1\sigma$-error bars are calculated from the jackknife method by equation (\ref{jackknife}) with the $N^\prime =7$.}
\label{fig4}
\end{figure}

A power-law model of the same form as equation (\ref{powerlaw}) has also been fitted to the measured $\xi(r)$ of GRBs above $100h^{-1}$ Mpc in Figure \ref{fig4}. The parameters $r_0$ and $\gamma$ are determined by a least-$\chi^2$ procedure. In the $\Lambda$CDM cosmology, we find a best-fit power law with $r_0= (1235.2 \pm 342.6)~h^{-1}$ Mpc and $\gamma = 0.80\pm 0.19 $ ($1\sigma$ confidence level) on scales $r=200$ to $10^4~h^{-1}$ Mpc. In the EdS Universe, the results are $r_0= (322.4 \pm 92.3)~h^{-1}$ Mpc and $\gamma = 0.62\pm 0.20 $ ($1\sigma$ confidence level).
The values of $r_0$ and $\gamma$ for both cosmological models are presented in Table \ref{table3}. The scale-length $r_0$ in the power law $\xi(r)=(r/r_0)^{-\gamma}$ is not very well determined in the analysis. That is because the data points of measured $\xi$ have large error bars, due to the limited size of the GRB sample. This would be improved once a larger sample of GRBs with well-determined redshifts are available for the analysis.

\begin{figure}
\includegraphics[scale=0.55]{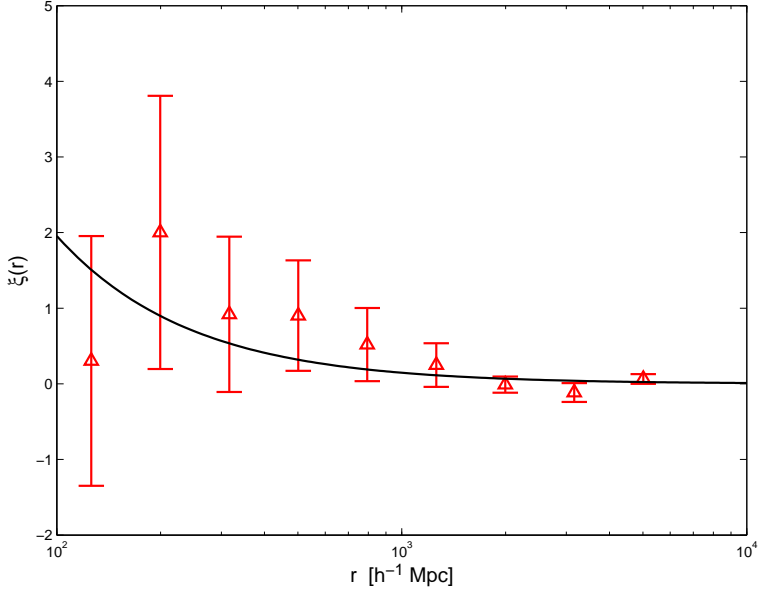}
\caption{The best-fit power law of the measured $\xi(r)$ for the GRBs at $z<1.5$ from \citet{GRBs} for the $\Lambda$CDM universe with $(\Omega_{{\rm m}}, \Omega_{{\rm \Lambda}})=(0.28,0.72)$. Triangles represent the measured $\xi(r)$ obtained from that the density of random points is 20 times of that of the GRB data. The $1\sigma$-error bars are calculated from the jackknife method by equation (\ref{jackknife}) with the $N^\prime =5$. Solid black lines indicate the best-fit power law of $\xi(r)$. The best-fit values of the correlation length $r_0$ and the slope $\gamma$ are $r_0=181.4 \pm 113.9$ $h^{-1}$ Mpc and $\gamma =1.12\pm 0.46$, with ${\bar \chi^2}_{{\rm min}}=0.54$.}
\label{fig5}
\end{figure}

Most galaxies have a redshift $z<1.5$. For a comparison with the best-fit power law to that of galaxies, we fit a power law to those GRBs with $z<1.5$. This sub-catalogue contains 173 objects. We obtained a best-fit $r_0=181.4 \pm 113.9$ $h^{-1}$ Mpc and $\gamma =1.12\pm 0.46$, with ${\bar \chi^2}_{{\rm min}}=0.54$. The best-fit result was presented in Figure \ref{fig5}. The $1\sigma$-error bars are calculated from the jackknife method by equation (\ref{jackknife}) with the $N^\prime =5$.

\section{Conclusions and Discussions}\label{conclusion}
Up to now, the studies of the correlation function $\xi(r)$ have usually been limited to the galaxy samples at low redshift, i.e. $0<z \lesssim 1$.  
In this paper, we extended this work to the GRBs samples. Many of them have a redshift range of $z>1$ and thus can be used to explore the matter clustering and its evolution in the earlier Universe. We calculated the 2PCF $\xi(r)$ for the GRB samples from \citet{GRBs}.
We considered two popular cosmological model in estimating $\xi(r)$,  a $\Lambda$CDM cosmology with $(\Omega_{{\rm m}}, \Omega_{{\rm \Lambda}})=(0.28,0.72)$, and an EdS one with $(\Omega_{{\rm m}}, \Omega_{{\rm \Lambda}})=(1,0)$. We found that in the $\Lambda$CDM Universe, the $\xi(r)$ for the GRB samples on large $r$ ($\simeq 200$ to $10^4~h^{-1}$ Mpc) follows a best-fit power law with $r_0= (1235.2 \pm 342.6)~h^{-1}$ Mpc and $\gamma = 0.80\pm 0.19 $ ($1\sigma$ confidence level). For the EdS Universe, the results are $r_0= (322.4 \pm 92.3)~h^{-1}$ Mpc and $\gamma = 0.62\pm 0.20 $ ($1\sigma$ confidence level). Both correlation functions showed systematic deviations from the best-fit power law below the scale $r=200h^{-1}$ Mpc. We concluded these departures are due to the volume-limited sample, which has insufficient GRB pairs with separation below $100h^{-1}$ Mpc. In fact, on the scale  $r< 100h^{-1}$ Mpc, the matter distribution in our Universe is usually determined by the redshift galaxy survey, such as the Sloan Digital Sky Survey (SDSS) \citep{SDSS}. Our work provided a supplement for the measurement of the matter correlation on scales $r> 200h^{-1}$ Mpc. For a better performance of the measurement of $\xi(r)$ for GRBs on smaller scales, which can be used to compare with the results obtained from the galaxy survey below $100h^{-1}$ Mpc, further observations are needed.

In fact, since the GRB sample we used is not large enough, one has to take these results with a grain of salt. Besides, the Wilkinson Microwave Anisotropy Probe (WMAP)  \citep{Bennett2013} and the PLANCK satellite \citep{Planck2013} have provided unprecedentedly precise measurements of the anisotropy and the inhomogeneity of the matter distribution in the early Universe. To answer the questions mentioned above, one may have to use the WMAP/PLANCK observational data for a combined analysis. A more careful clustering analysis of the medium redshift Universe using the WMAP/PLANCK observations  together with the quasars and GRB data is currently undertaking. We hope that the results would shed new light on the structure formation theory and the inhomogeneities of our early Universe.

\acknowledgments
We are grateful to Zhi-Bing Li from the Sun Yat-Sen University and Zhe Chang from the Institute of High Energy Physics for all the useful discussions and suggestions. The final version of the manuscript has greatly benefited from the careful scrutiny of the referee. This work is based on the GRB catalogue presented by Jochen Greiner at \url{http://www.mpe.mpg.de/~jcg/grbgen.html}. 






\clearpage

\centering

\centering

\centering

\centering

\centering

\clearpage

\begin{table}
\begin{center}
\caption{Properties of the GRB catalogue presented by \citet{GRBs}: the redshift bins and the number of GRBs in each of the $N_b=7$ bins that are used to calculate the `FtF' error.\label{table1}}
$~$\\
\begin{tabular}{crrr}
\tableline\tableline
Redshift & Number &
\\ 
\tableline
$0<z\leq1$ & 112 & \\
$1<z\leq2$ & 108 & \\
$2<z\leq3$ & 80 & \\
$3<z\leq4$ & 40 & \\
$4<z\leq5$ & 21 & \\
$5<z\leq6$ & 8 & \\
$6<z\leq7$ & 4 & \\
\tableline
\end{tabular}
\end{center}
\end{table}


\begin{table}
\begin{center}
\caption{The power-law fit to the measured $\xi(r)$ of the GRBs catalogue presented by \citet{GRBs}. The best-fit values of the correlation length $r_0$ and the slope $\gamma$ for a $\Lambda$CDM universe with $(\Omega_{{\rm m}}, \Omega_{{\rm \Lambda}})=(0.28,0.72)$ and an EdS universe with $(\Omega_{{\rm m}}, \Omega_{{\rm \Lambda}})=(1,0)$ over the scales $0<r\leq 800~h^{-1}$ Mpc are presented. ${\bar \chi^2}_{{\rm min}}=\chi^2_{{\rm min}}/(N-d.o.f+1)$ represents the reduced minimal chi-squares of the results on the scale $r>100h^{-1}$ Mpc. $N=9$ is the number of the bins. $d.o.f=2$ represents the degree of freedom, i.e. the number of free parameters in the fit. \label{table3}}
$~$\\
\begin{tabular}{crrr}
\tableline\tableline
 & $\Lambda$CDM & EdS
\\ 
\tableline
$r_0$ & $1235.2 \pm 342.6$ & $322.4\pm92.3$ \\
$\gamma$ & $0.80\pm 0.19$ & $0.62\pm0.20$ \\
${\bar \chi^2}_{{\rm min}}$ & $0.19$ & $0.21$ \\
\tableline
\end{tabular}
\end{center}
\end{table}



\end{document}